\documentclass[a4paper]{article}

\usepackage[dvips]{graphicx}

\begin{document}

\pagestyle{empty}

\begin{center}

\large
\bf{Korean VLBI Network:\\
the First Dedicated Mm-Wavelength VLBI Network in East Asia}

\bigskip

\normalsize
\bf{\underline{Kiyoaki Wajima}, Hyun-Goo Kim, Seog-Tae Han, Duk-Gyoo Roh, \\
Do-Heung Je, Se-Jin Oh, Seog-Oh Wi, and Korean VLBI Network Group}

\bigskip

\bf{\it Korea Astronomy and Space Science Institute, 61--1 Hwaam-dong,
Yuseong, Daejeon 305--348, KOREA \\
E-mail: kiyoaki@trao.re.kr}

\end{center}

\begin{abstract}
Korean VLBI Network (KVN) is the first dedicated mm-wavelength VLBI Network
in East Asia and will be available from the middle of 2008.
KVN consists of three stations and has the maximum observation frequency
of 129 GHz with the maximum baseline length of 480 km.
KVN has unique characteristics in the multifrequency, simultaneous observing
system.
By taking advantage of this we are considering various science topics,
including not only maser
emitting regions and young stellar objects in our galaxy, but also
extragalactic objects.
Construction of the first site is in progress.
We are concurrently developing components, including receivers, data
acquisition systems, and a correlator, and also arranging the international
collaboration.
\end{abstract}

\bigskip
\bigskip
\noindent
{\bf INTRODUCTION}
\bigskip

\begin{figure}[htbp]
\centering
\includegraphics[width=0.95\linewidth]{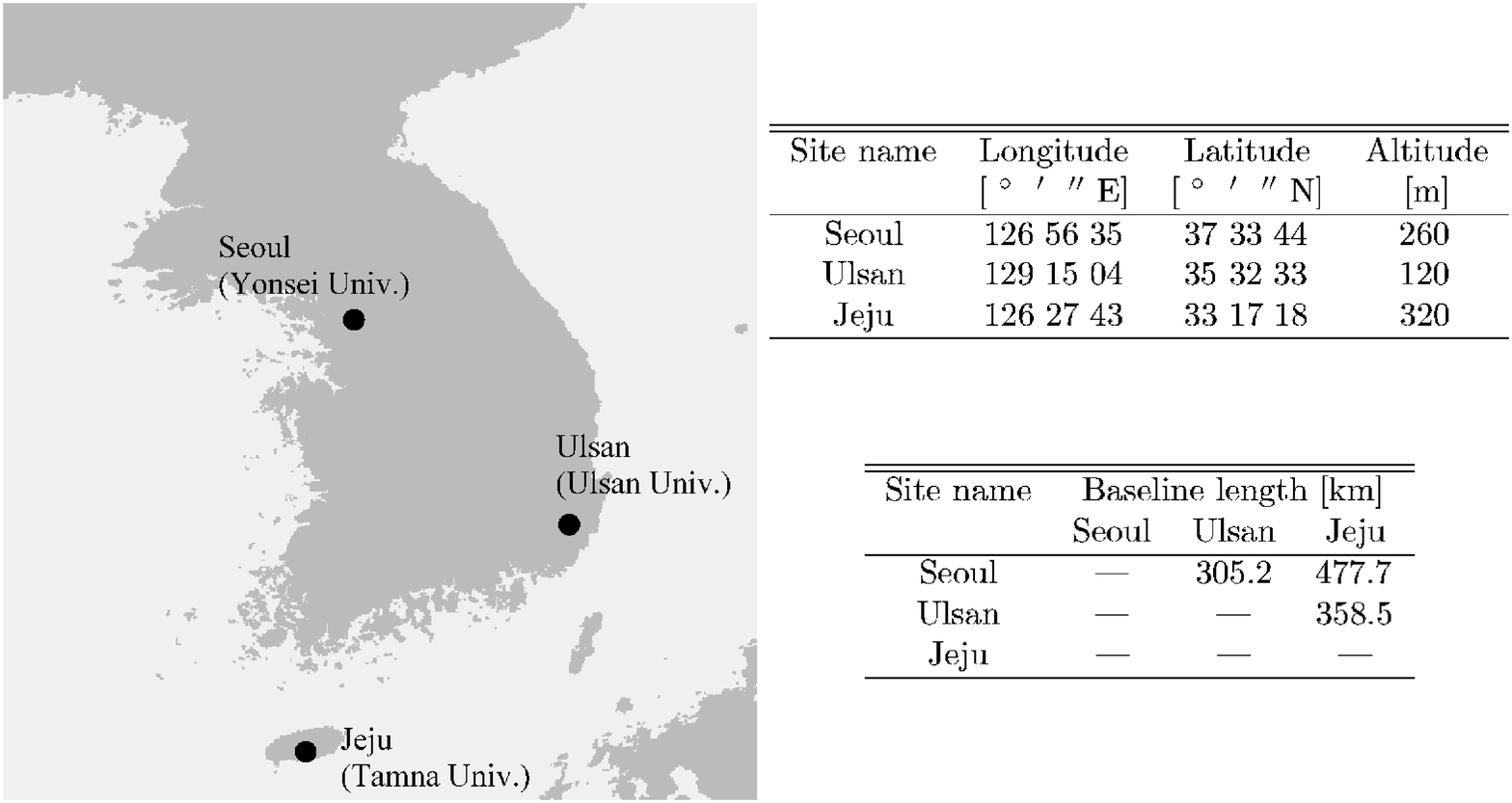}
\caption{Location of KVN sites. Site position paramters and baseline length
between each station are also listed.}
\label{fig:KVN Sites}
\end{figure}

KVN is the first VLBI facility in Korea [1].
KVN consists of three stations, Seoul (Yonsei Univ.), Ulsan (Ulsan Univ.),
and Jeju (Tamna Univ.) and has the maximum baseline length of 480\,km
(Seoul -- Jeju) with the maximum observation frequency of 129\,GHz
(see Fig.\ \ref{fig:KVN Sites}).
The maximum angular resolution of KVN is therefore 1 milliarcsecond (mas).
This frequency is the highest one among dedicated VLBI facilities around
the world.

The antenna has a diameter of 21\,m and can perform a
fast-switching (3 deg sec$^{-1}$ in Az, El)
to enable phase-referencing VLBI observations and superposition of different
frequency images.

\begin{figure}[htbp]
\centering
\includegraphics[width=0.75\linewidth]{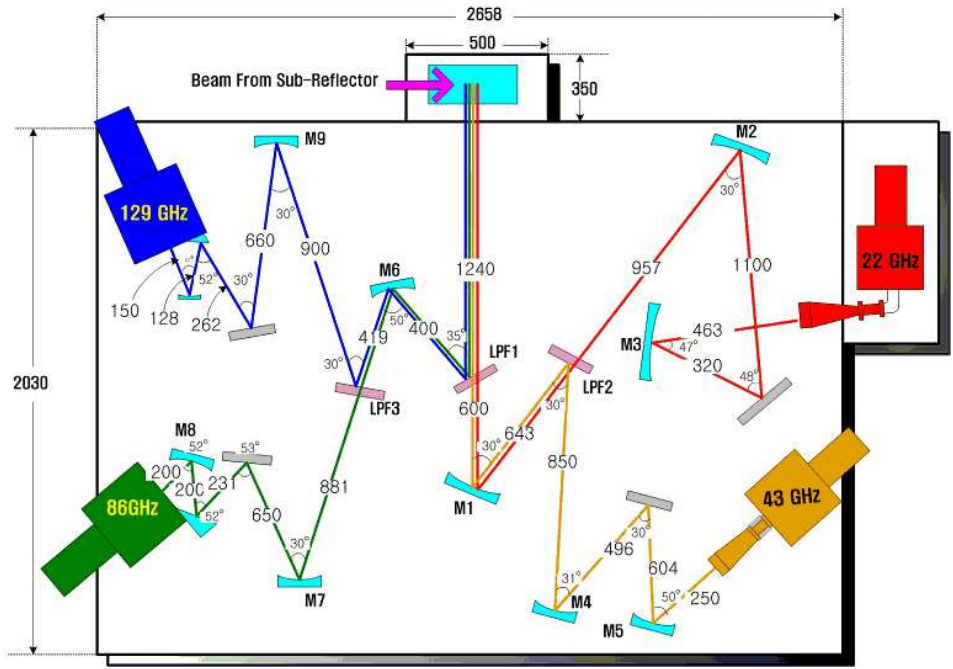}
\caption{Quasi-optics system of KVN.}
\label{fig:QuasiOpt}
\end{figure}

KVN employs unique observing system.
KVN has simultaneous reception system up to four frequency channels,
22, 43, 86, and 129\,GHz.
In order to ensure this, three frequency selective surfaces (FSSs) are
used in the quasi-optics system (see Fig.\ \ref{fig:QuasiOpt}).
Dual frequency receiver at 2 and 8\,GHz is also equipped for geodetic
observations.
Dual polarization reception will be available at whole observation
frequencies.

KVN will be constructed with two-term planning.
In the first stage (2001 -- 2007), fundamental construction of each site,
installation of the antenna and the data acquisition system, and
receivers for 22 and 43\,GHz will be carried out and test observations
will be performed in order to estimate the capabilities of the whole
observing system at lower frequencies.
Receivers at 86 and 129\,GHz will be installed in the second stage
(2008 -- 2010).
KVN correlator project is also ongoing concurrently with the KVN project
with 5-year mission since 2004.

\bigskip
\bigskip
\noindent
{\bf SCIENTIFIC TARGETS}
\bigskip

By taking advantage of the unique system, mentioned in the previous section,
we are considering various scientific targets.
In evolved stars and star forming regions, which are main targets in KVN
observations, we intend to study physical relations between formation and
pulsation driven shocks and SiO pumping mechanisms by observing H$_2$O and
SiO masers simultaneously.
KVN covers three SiO maser lines with different rotational transitions,
simultaneous observation of these emissions is therefore essential to
study pumping mechanism of SiO maser.
Young stellar objects (YSOs) will also be a key target in order to study
magnetosphere structure by monitoring multifrequency flux densities to
detect flare features.

Simultaneous reception system will also be a good one to reveal parsec-scale
accretion processes of warm gas in active galactic nuclei (AGNs).
In nearby galaxies, KVN will be able to distinguish free-free absorption
from synchrotron self absorption in the jets of AGN.
H$_2$O megamasers are also important targets and the position of maser
spots will be precisely determined with respect to a continuum emission by
using KVN.

Not only various scientific targets but also geodetic VLBI observations
will be performed on the international collaboration.

\clearpage
\noindent
{\bf SITE CONSTRUCTION AND DEVELOPMENT OF COMPONENTS}
\bigskip

Fundamental construction of the first site (Ulsan) started from the last
December and is in progress.
The second (Jeju) and third (Seoul) sites will be started construction
by the end of 2005 and the middle of 2006, respectively.
We have already made a contract and completed the detailed design of the
antenna.
The antenna installation at Ulsan site will start from the early 2006
and completion of installation at all the three sites will be in the
middle of 2007.

\begin{figure}[tbp]
\centering
\includegraphics[width=0.7\linewidth]{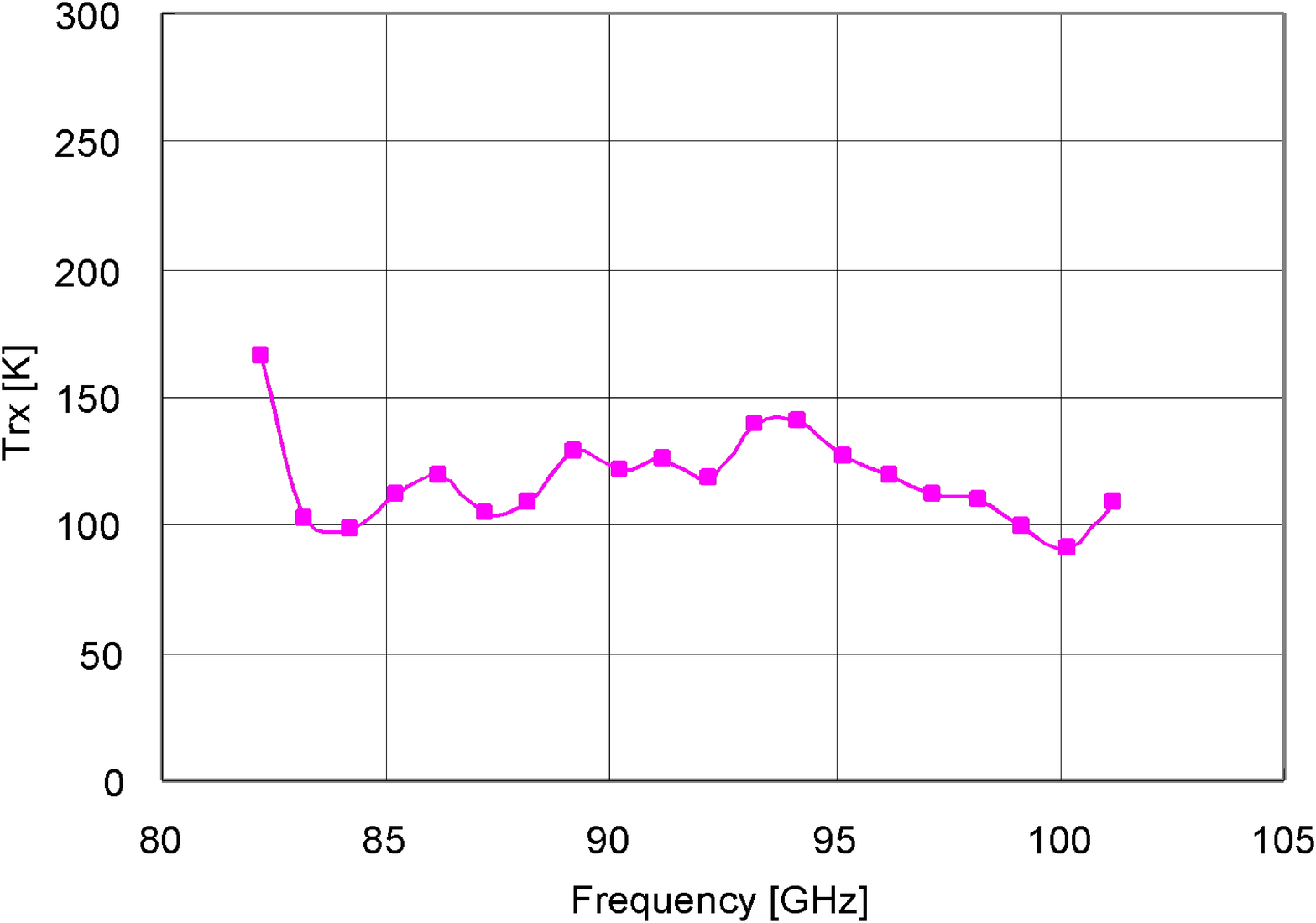}
\caption{Measurement result of the receiver noise temperature at 86\,GHz
developed by KASI.}
\label{fig:KVN_RX_86}
\end{figure}

Development of MMIC HEMT amplifiers is in progress for observations at
22, 43, and 86\,GHz.
The receiver noise temperature ($T_{\rm rx}$) for each frequency is typically
65\,K (43\,GHz) and 120\,K (86\,GHz) and those have sufficient performance
as expected (see Fig.\ \ref{fig:KVN_RX_86}).
We will perform further improvement of those amplifiers, and start development
of an SIS mixer for 129\,GHz observations in 2007.

A design of the data acquisition system (DAS) has already completed.
Four 1-Gsps samplers are equipped and observed data at each frequency
are digitized in the data rate of 2\,Gbps.
A digital filter is installed in order to accommodate various observation
modes by simultaneous receiving system.
The data are recorded by the Mark 5B recorder developed at MIT Haystack
Observatory [2].
Hardware connection and data transmission of almost all DAS equipments are
compliant with the VLBI Hardware Standard Interface (VSI-H).

The KVN correlator project is ongoing and we are currently studying
required functions and how to carry out the program.
We plan optical fiber connection of each KVN site in the future in order
to realize e-VLBI observations, and we are discussing the possibility
with domestic communication companies.

\bigskip
\bigskip
\noindent
{\bf INTERNATIONAL COLLABORATIONS}
\bigskip

As stated in the first section, we have already started international
collaboration in parallel with the site construction of KVN.

In 2002 VLBI collaboration agreement was concluded between Korea Astronomy
and Space Science Institute (KASI, former Korea Astronomy Observatory) and
National Astronomical Observatory of Japan (NAOJ).
On the basis of the agreement,
we have already started discussion on the future collaboration for joint
observations between KVN and Japanese VLBI network, VERA (VLBI Exploration
of Radio Astrometry), and on science and technology exchange.
We have already succeeded 86\,GHz VLBI observations between Taeduk 14\,m --
Nobeyama 45\,m baseline toward SiO maser emission in VY CMa [3].
This is the first detection of SiO maser emission at 86\,GHz toward this
source.
We are planning further VLBI observations for drawing up future joint
observations.

The standing committee for the coordination of the East Asia VLBI Consortium,
including Korea, Japan, and China, was established in order to discuss the
future collaboration and to organize the East Asia VLBI network (EAVN) [4].
EAVN will be organized up to 16 radio telescopes in each country, and
KVN and VERA will become core stations in this array.

KASI and NAOJ have made an arrangement for development of the VLBI correlator.
The correlator is mainly designed for joint observations with KVN and VERA,
while current correlator design involves a sufficient number of stations and
wide delay tracking window so that correlation for observations
of EAVN and the next Space VLBI mission, mentioned below, can be performed.
The correlator will be installed at the KVN site in Seoul and completed on
2008.

We are also considering to participate the next Space VLBI (VSOP-2) mission
[5], [6] as a ground counterpart and data processing work by a correlator
above mentioned.
KVN and VERA will be one of the important ground stations for the VSOP-2
mission because all radio telescopes including VSOP-2 spacecraft furnish with
the dual polarization receiving system at 22 and 43\,GHz and the
fast-switching function.

\bigskip
\bigskip
\noindent
{\bf CONCLUDING REMARKS}
\bigskip

KVN is the first dedicated mm-VLBI facility with an observing frequency up to
129\,GHz.
The KVN project is well under way thanks to collaboration with a lot of
institutes and companies around the world.
It is our wish to make many observations not only by KVN but also with VLBI
facilities around the world and produce many scientific results.
You can find the up-to-date information on the KVN and related projects via
Internet (http://www.trao.re.kr/{\symbol{'176}}kvn/).

\bigskip
\bigskip
\noindent
{\bf ACKNOWLEDGMENTS}
\bigskip

K.W.\ acknowledges support from Korea Science and Engineering Foundation
(KOSEF).

\bigskip
\bigskip
\noindent
{\bf REFERENCES}
\bigskip

\noindent
[1] Y.~C.~Minh, D.~-G.~Roh, S.~-T.~Han, \& H.~-G.~Kim, ``Construction of the
Korean VLBI Network (KVN),''
\hspace*{11pt}
in {\it ASP Conf.\ Ser.\ vol.\ 306, New
Technologies in VLBI}, ed.\ Y.~C.~Minh (San Francisco: ASP), pp.\ 373--381,
\hspace*{11pt}
2003.

\noindent
[2] A.~Whitney, `` The Mark 5B VLBI Data System,'' in {\it Proc.\ 7th EVN
Symp.}, pp.\ 251--252, 2004.

\noindent
[3] K.~M.~Shibata et al., ``First mm-VLBI Observations between the TRAO
14-m and the NRO 45-m Telescopes:
\hspace*{11pt}
Observations of 86\,GHz SiO Masers in VY Canis Majoris,''
{\it Publ. Astron. Soc. Japan}, vol.\ 56, pp.\ 475--480,
\hspace*{11pt}
2004.

\noindent
[4] M.~Inoue, ``East Asia VLBI Consortium and Its Committee,''
{\it J.\ Korean Astron.\ Soc.}, vol.\ 38, pp.\ 77--79,
\hspace*{11pt}
2005.

\noindent
[5] H.~Hirabayashi et al., ``The VSOP-2 Project: A Second Generation
Space-VLBI Mission Ranging to mm-
\hspace*{11pt}
Wavelengths,'' in {\it Proc.\ SPIE Int.\ Soc.\
Opt.\ Eng.}, vol.\ 5487, pp.\ 1646--1656, 2004.

\noindent
[6] Y.~Murata, ``Space VLBI Project,''
{\it J.\ Korean Astron.\ Soc.}, vol.\ 38, pp.\ 97--100, 2005.

\end{document}